\documentstyle[sprocl,epsfig]{article}

\def\Journal#1#2#3#4{{#1} {\bf #2}, #3 (#4)}

\def\NIMA{{\em Nucl. Instrum. Methods} A}
\def\NIMB{{\em Nucl. Instrum. Methods} B}

\def\NPBP{{\em Nucl. Phys.} B (Proc. Suppl.)}

\def\PLB{{\em Phys. Lett.}  B} 
\def\PRL{\em Phys. Rev. Lett.} 
\def\PRD{{\em Phys. Rev.} D}

\def\PRP{{\em Phys. Rep. }}
\def\EPC{{\em Europ. Phys. J.} C}

\def\PAN{\em Phys. Atom. Nucl.}

\def\APJ{\em ApJ}

\def\ASP{\em Astroparticle Physics}

\def\IJMP{\em Int. Journal of Modern Physics}

\def\ra{\rightarrow}

\def\be{\begin{equation}}
\def\ee{\end{equation}}
\def\bea{\begin{eqnarray}}
\def\eea{\end{eqnarray}}

\newcommand{\expe}{experiment }

\newcommand{\exps}{experiments }

\newcommand{\majo}{Majorana }

\newcommand{\osz}{oscillation }

\newcommand{\delm}{\mbox{$\Delta m^2$} }

\newcommand{\me}{\mbox{$m_{\nu_e}$} }

\newcommand{\mmu}{\mbox{$m_{\nu_\mu}$} }
\newcommand{\mtau}{\mbox{$m_{\nu_\tau}$} }

\newcommand{\nel}{\mbox{$\nu_e$} }
\newcommand{\nmu}{\mbox{$\nu_\mu$} }
\newcommand{\ntau}{\mbox{$\nu_\tau$} }
\newcommand{\sint}{\mbox{$sin^2 2\theta$} }

\newcommand{\munu}{\mbox{$\mu_{\nu}$} }
\newcommand{\mub}{\mbox{$\mu_B$} }

\newcommand{\mamo}{magnetic moment}

\newcommand{\neu}{neutrino }
\newcommand{\neus}{neutrinos}

\newcommand{\rehsa}{\mbox{$^{187}Re$ }}

\newcommand{\cref}{\mbox{$^{51}Cr$ }}

\begin{document}

\title{Non-accelerator neutrino mass searches}

\author{K. Zuber}

\address{Lehrstuhl f\"ur Exp. Physik IV, Universit\"at Dortmund,\\
44221 Dortmund, Germany}



\maketitle\abstracts{The current status of non-accelerator based searches for
effects of a non-vanishing neutrino mass is
reviewed. Beside the direct kinematical methods this includes
searches for magnetic moments and a discussion of the solar
neutrino problem.}
\section{Introduction}
Neutrinos play a fundamental role in several fields of physics from cosmology down to
particle physics. Even more, the observation of a non-vanishing rest mass of neutrinos
would
have a big impact on our present model of particle physics and might guide towards grand
unified
theories. Currently three evidences exist showing effects of massive neutrinos: the
deficit in
solar neutrinos, the zenith angle dependence of atmospheric neutrinos and the excess
events
observed by LSND. These effects are explained with the help of neutrino
oscillations, thus depending on \delm{} = $m_2^2 - m_1^2$, where $m_1,m_2$ are the
\neu{} mass
eigenvalues and therefore are not absolute mass measurements. 
For a recent review on the physics of massive neutrinos see \cite{zub98}.
\section{Mass measurements of the electron neutrino}
The classical way to determine the mass of $\bar{\nel}$ (which is identical to $m_{\nu_e}$
assuming CPT invariance) is the
investigation of the
electron spectrum in beta decay.
A finite \neu mass will reduce the phase space and leads to a 
change of the shape
of the electron spectra.
In case several mass
eigenstates contribute, the total electron spectrum is given by a 
superposition
of the individual
contributions
\be
N(E) \propto F(E,Z) \cdot p \cdot E \cdot (Q-E) \cdot \sum^3_{i=1} 
\sqrt{(Q-E)^2 - m_i^2}
\mid U_{ei}^2
\mid 
\ee
where F(E,Z) is the Fermi-function, $m_i$ are the mass eigenvalues, $U_{ei}^2$ are
the mixing matrix elements connecting weak and mass eigenstates and $E,p$ are energy and momentum
of the emitted electron. The different involved $m_i$ produce kinks 
in the Kurie-plot 
where the size of the kinks is a measure
for the corresponding mixing
angle. 
Searches for an eV-\neu are done near the endpoint region of isotopes with low Q - values.
The preferred isotope under study is tritium, with an endpoint energy of about 18.6 keV.
\begin{figure}
\begin{center}
\epsfig{file=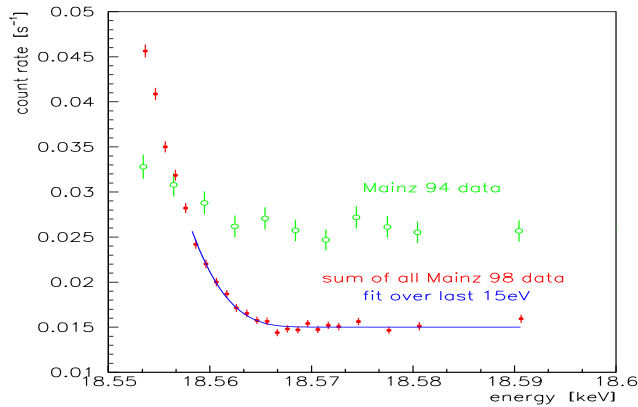,width=9cm,height=6cm} 
\caption{Mainz 1998 electron spectrum near the endpoint of tritium decay. The
signal/background ratio is increased by a factor 
of 10 in comparison with the
1994 data. The Q-value of 18.574 keV is marking to the center of mass of the rotation-vibration 
excitations of the
molecular ground state of the daughter ion $^3HeT^+$.}
\label{pic:mainz}
\end{center}
\end{figure}
The currently running \exps in Mainz and Troitzk are using electrostatic retarding
spectrometers \cite{lob85,pic92}. 
Fig.\ref{pic:mainz} shows the
current electron spectrum near the
endpoint as obtained with the Mainz spectrometer. 
The current obtained limits are 2.8 eV (95 \% CL) ($m_{\nu}^2 = - 3.7 \pm 5.3 (stat.) \pm 2.1 (sys.) eV^2$)
\cite{wei99} and 2.5 eV (95 \% CL)
($m_{\nu}^2 = - 1.9 \pm 3.4 (stat.) \pm 2.2 (sys.) eV^2$)
\cite{lob99} respectively. The final sensitivity should be around 2 eV.\\
Beside this, the Troitzk \expe observed excess counts in the
region of interest,
which can be described by a monoenergetic line a few eV below the endpoint. 
Even more, a semiannual modulation of the line position is observed \cite{lob99}. Clearly
further 
measurements are needed to investigate this effect. Considerations of building a new larger scale
version of such a spectrometer exist to probe neutrino masses down below 1 eV.\\
A complementary strategy is followed by using cryogenic microcalorimeters. Because these
\exps measure the total energy released,
final state effects are not important. This method allows the investigation
of the $\beta$-decay of \rehsa, which has the lowest Q-value of all $\beta$-emitters (Q=2.67
keV). Furthermore the associated half-life measurement would be quite important, because
the \rehsa - $^{187}$Os pair is a well known cosmochronometer and a more precise half -
life measurement would sharpen the dating of events in the early universe like the formation of 
the solar system.
Cryogenic bolometers were build in form of metallic Re as well as AgReO$_4$ crystals and
$\beta$ - spectra 
were measured \cite{gat99} \cite{ale99}, but at present 
the experiments
are not
giving any limits on \neu masses. Investigations to use this kind of technique also for
calorimetric measurements on tritium \cite{dep99} and on $^{163}$Ho \cite{meu98} are currently done.
Measuring accurately
branching ratios of atomic transitions or the internal bremsstrahlung spectrum in $^{163}$Ho
is interesting because this would result directly in a limit on \me.

\section{Mass measurement of the muon \neu{}}
The way to obtain limits on \mmu is given by the two-body decay of
the $\pi^+$.
A precise measurement of the muon momentum $p_{\mu}$ and
knowledge of $m_{\mu}$
and
$m_{\pi}$ is required. 
This measurement was done at the PSI resulting in a limit of \cite{ass96}
\be
\mmu^2 = (-0.016 \pm 0.023) MeV^2 \quad \ra \quad \mmu < 170 keV (90
\%CL)
\ee
A new idea looking for pion decay in flight using the g-2 storage ring at BNL has been proposed
recently \cite{cus99}. Because the g-2 ring would act as a high resolution spectrometer an
exploration
of \mmu down to 8 keV seems possible. Such a bound would have some far reaching consequences:
It would bring any magnetic moment calculated within
the 
standard model and associated with \nmu down to a
level of vanishing
astrophysical importance. Furthermore it would once and for all exclude that a possible 17 keV
mass eigenstate is the dominant contribution of \nmu . Possibly the largest impact is on
astrophysical topics. All bounds on \neu properties derived from stellar evolution are typically
valid for \neu masses below about 10 keV, so they would then apply for \nmu as well. For example, plasma
processes like $\gamma \ra \nu \bar{\nu}$ would contribute to stellar energy losses and significantly prohibit
helium ignition, unless the neutrino has a magnetic moment smaller than $\mu_{\nu} < 3 \cdot 10^{-12} \mu_B$
\cite{raf99} much more stringent than laboratory bounds.

\section{Mass measurement of the tau \neu{}}
The present knowledge of the mass of \ntau stems from measurements with
ARGUS, CLEO, OPAL, DELPHI and ALEPH (see \cite{pas97}).
Practically all \exps use the $\tau$-decay into five charged pions
$\tau \ra \ntau + 5\pi^{\pm} (\pi^0)$
To increase the
statistics CLEO, OPAL, DELPHI
and ALEPH
extended their search by including the 3 $\pi$ decay mode. But even with the 
disfavoured statistics,
the 5 prong-decay is much more sensitive, because the mass of the 
hadronic system peaks at about 1.6 
GeV, while the 3-prong system is dominated by the $a_1$ resonance at 
1.23 GeV. While ARGUS obtained their limit by investigating the invariant mass of the 
5 $\pi$-system, ALEPH, CLEO and OPAL 
performed a two-dimensional analysis by including the energy of 
the hadronic system.
The most 
stringent bound of \mtau $<$ 18.2 MeV is given by ALEPH \cite{bar98}.

\section{Magnetic moment of the neutrino}
Another possibility to check the \neu{} character and mass is the search for its
\mamo{}.
In the case of Dirac \neus{}, it can be shown that \neus{} can have a \mamo{}
due to loop diagrams which is proportional to
their mass and is given by \cite{lee77,mar77}
\be
\munu = \frac{3 G_F e}{8 \sqrt{2} \pi^2} m_{\nu} = 3.2 \cdot 10^{-19} (\frac{m_\nu}{eV}) \mub
\ee
In case of \neu masses in the eV-range, this is far to small to be observed
and to have any significant
effects in
astrophysics. Nevertheless
there exist GUT-models, which are able to increase the \mamo{} without increasing the
mass
\cite{pal92}. However
\majo{} \neus{} still have a vanishing static moment because of CPT-invariance.
The existence of diagonal terms in the \mamo{} matrix would therefore prove 
the
Dirac-character of \neus{}.
Non-diagonal terms in the moment matrix are possible for both types of \neus{}
allowing transition moments of the form \nel - $\bar{\nu}_\mu$.\\ 
Limits on magnetic moments arise from \nel $e$ - scattering \exps and 
astrophysical considerations. The 
differential cross section for \nel $e$ -
scattering in presence of a \mamo{} is given by
\be
\frac{d \sigma}{dT} = 
\sigma_{SM} + \frac{\pi \alpha^2 {\mu_{\nu}^2}}{m_e^2}
\frac{1-T/E_\nu}{T}
\ee
where $\sigma_{SM}$ is the standard model cross section and T is the kinetic energy of
the recoiling electron. 
As can be seen, the largest effect of a \mamo{} can be observed in the low
energy region, and because of
destructive interference
of the electroweak terms, searches with antineutrinos would be preferred.
Experiments done so far give limits of \munu{} $<1.8 \cdot 10^{-10} \mu_B$ (\nel), \munu $<7.4
\cdot
10^{-10} \mu_B$ (\nmu) and \munu{} $<5.4 \cdot 10^{-7} \mu_B$ (\ntau).
Astrophysical limits are somewhat more stringent but also more model dependent. 
To improve the experimental situation new experiments are taking data or are under construction. 
From the considerations mentioned before the obvious sources for searches
are nuclear reactors.
The most
advanced is the
MUNU \expe \cite{ams97} currently running at
the Bugey
reactor. It consists of a 1 m$^3$ TPC loaded with CF$_4$ under a pressure of 5 bar. The usage 
of a TPC will not only
allow to measure the electron energy but for the first time in such \exps also the 
scattering angle, making the
reconstruction of the neutrino energy possible. 
The expected
sensitivity level is down to $\munu = 3 \cdot 10^{-11} \mub$ . The usage 
of a low background Ge-NaI
spectrometer in a shallow depth near a reactor has
also been considered \cite{bed97}. Under investigation are also large low-level detectors
with a low-energy threshold
of a few keV in underground laboratories. The reactor 
would be replaced
by a strong $\beta$-source. 
Calculations for a scenario of a 1-5 MCi $^{147}$Pm 
source (endpoint
energy of 234.7 keV) in combination with a 100 kg low-level NaI(Tl) detector with a 
threshold of about 2
keV
can be found in \cite{bar96}. Also using a $^{51}$Cr source within the BOREXINO experiment will allow
to put stringent limits on $\munu$. 
\begin{figure}[hh]
\begin{center}
\begin{tabular}{ll}
\mbox{\epsfig{file=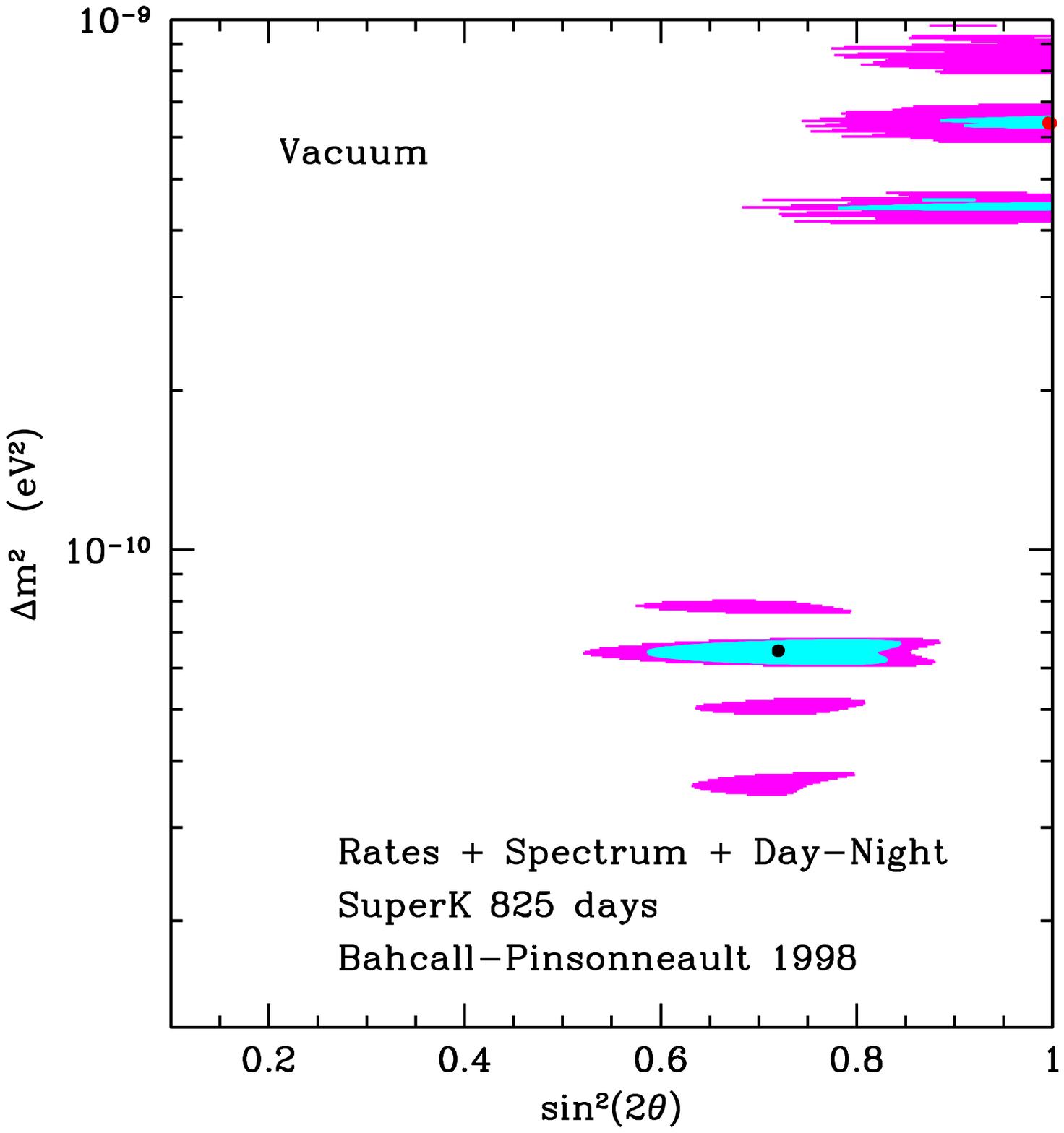,height=7cm,width=5.6cm}} &
\mbox{\epsfig{file=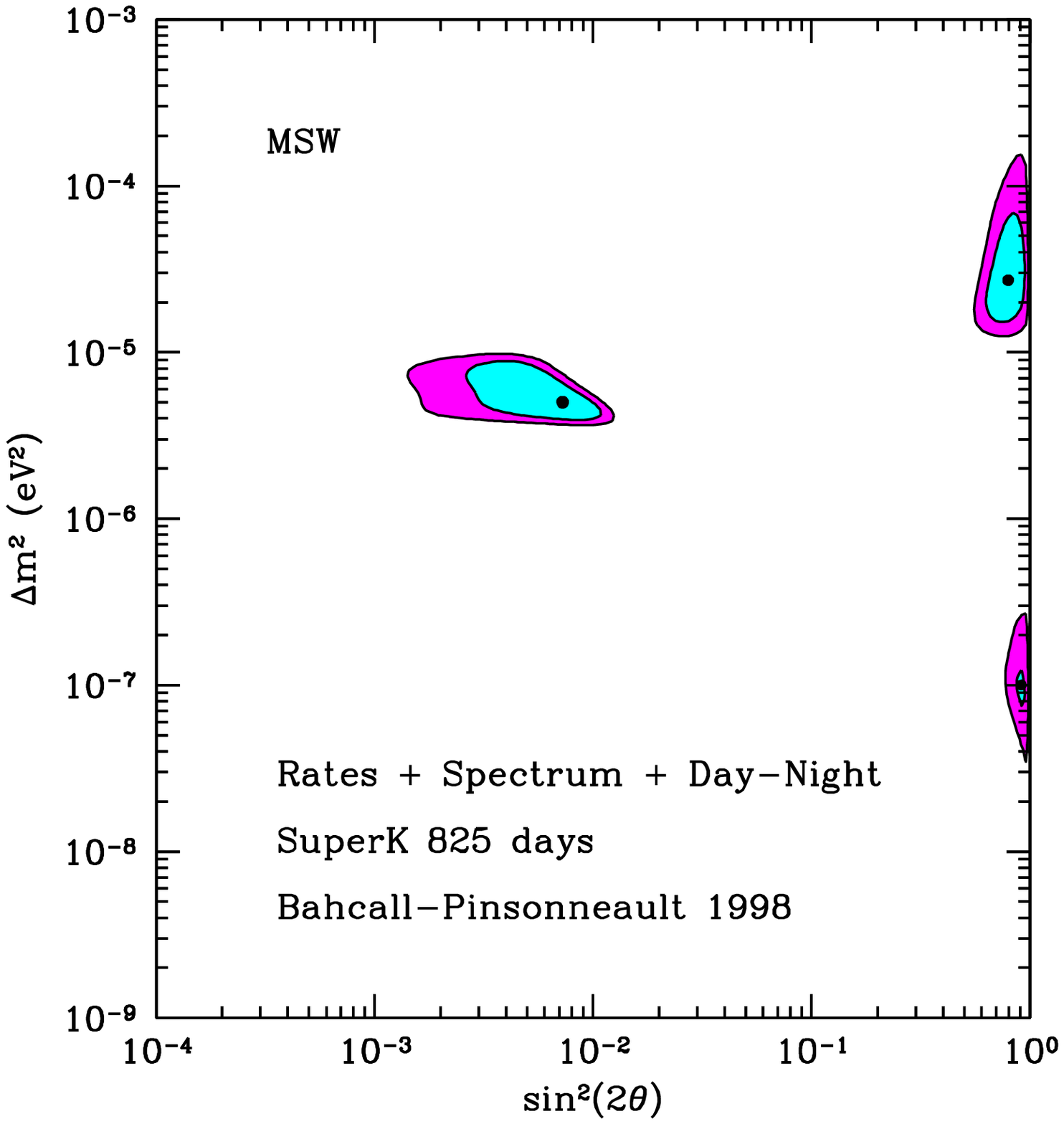,height=7cm,width=5.6cm}}
\end{tabular}
\caption{Left: Current solutions to the solar neutrino problem in terms of
neutrino
oscillations. Shown are the Vacuum oscillations (VO) at \delm around
$10^{-10} eV$
and a large mixing angle \sint.
Right: The MSW solutions at around \delm $ \approx 10^{-5} eV^2$ and \sint
$\approx$ 1
(Large
Mixing Angle, LMA) and around
$10^{-2} eV^2$ (Small Mixing Angle, SMA). Another solution with \sint
$\approx$ 1 and
\delm
$\approx 10^{-7} eV^2$ also seems possible.}
\label{pic:jnb}
\end{center}
\end{figure}

\section{Solar neutrinos}
An understanding of the sun is fundamental for the theory of stellar evolution, because
the sun is the only star we really can discuss and observe in great detail.
One aspect of
solar physics is the understanding of energy generation and the solar
interior, probed by the observation of solar \neus{}.
They cover
an energy range from keV up to about 15 MeV, where the overwhelming flux is due
to the pp-neutrinos, having an energy less than 430 keV. The most energetic
\neus{} come from $^8$B decay and the hep-neutrinos. 
The current status is shown in Tab. \ref{tab:exp}. 
Combining all results seems to suggest new \neu{} properties as the solution.
Explanations within
the framework of neutrino oscillations (Fig.\ref{pic:jnb})
include vacuum solutions (VO) as well as matter oscillations via the MSW-effect.
Fortunately the solutions disturb the solar \neu{} spectrum
in different ways allowing for experimental decisions.
Beside measuring the distortion in the energy spectrum, this includes day-night
and seasonal effects.
\begin{table}[ht]
\caption{Experimental status of solar neutrino experiments and theoretical
expectation using the standard solar model of \protect \cite{bah98}
\label{tab:exp}}
\vspace{0.2cm}
\begin{center}
\footnotesize
\begin{tabular}{|c|c|c|}
\hline
{} &\raisebox{0pt}[13pt][7pt]{Data } &
\raisebox{0pt}[13pt][7pt]{Theory}\\
\hline
{\raisebox{0pt}[12pt][6pt]{GALLEX}} & 77.5 $\pm 6.2 ^{+4.3}_{-4.7}$ SNU &
$129^{+8}_{-6}$ SNU\\
\hline
{\raisebox{0pt}[12pt][6pt]{SAGE}} & $66.9 \pm 8.9$ SNU &
$129^{+8}_{-6}$
SNU\\
\hline
{\raisebox{0pt}[12pt][6pt]{Homestake}} & 2.56 $\pm 0.16 \pm 0.15$ SNU &
$7.7^{+1.2}_{-1.0}$ SNU\\
\hline
{\raisebox{0pt}[12pt][6pt]{Super-K}} & $2.45 \pm 0.04
\pm 0.07^{+0.4}
_{-0.3} \cdot 10^6 cm^{-2}s^{-1}$ & $5.4 \pm 1.0 \cdot 10^6 cm^{-2}s^{-1}$\\
\hline
\end{tabular}
\end{center}
\end{table}
\\
The next step in clarifying the situation will be done by SNO, using 1 kt of $D_2$O
instead of
normal water \cite{bog99}.
They will measure three different reactions
\bea
\nel + D \ra p + p + e \quad (CC)\\
\nu_X + D \ra \nu_X + p + n \quad (NC)\\
\nel + e \ra \nel + e
\eea
By investigating the CC reaction SNO will be able to measure the $^8$B spectral shape
and by
comparing the flavour-sensitive CC with the flavour-blind NC reaction they will test the
oscillation
scenarios. The \expe{} started data taking recently and first results are expected
soon.
To measure \neus{} at lower energies in real time, you have to choose a different
detection
technique.
The next to come up is BOREXINO \cite{borex}, a 300 ton liquid scintillator
currently installed at Gran Sasso Laboratory. It is especially designed to measure the
$^7$Be \neus{}. Furthermore
proposals exist to measure even pp-neutrinos in real time in form of LENS
\cite{rag97}(100 t liquid
scintillator
containing a large amount of the double beta
isotope $^{176}$Yb, using an excited intermediate state for a coincidence measurement),
HELLAZ
\cite{tao97} (a 2000 m$^3$ high
pressure helium TPC at LN2 temperature using $\nu$-e scattering), HERON
\cite{ban95}(Liquid Helium
based
experiment using roton excitations generated by energy deposition in the
helium for detection)  and
SUPER-MUNU \cite{viu99} (high pressure TPCs - modular design - filled with CF$_4$ using
$\nu$-e   
scattering). All
this will finally
(hopefully)
give the full information on solar neutrinos and settle the question whether \neu{}
\osz{}
are responsible or not. 
Until such \exps{} will show up the measurement of pp-neutrinos
will 
continue with SAGE and GNO (a continuation and possible upgrade of GALLEX).
\smallskip\\
\begin{figure}[hhh]
\begin{center}
\epsfig{file=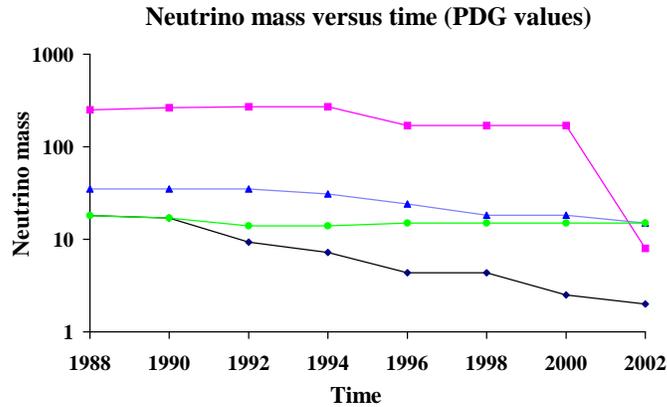,width=9cm,height=6cm}
\caption{Evolution of neutrino mass limits over the last 15 years using the Particle Data
Group values.
Extrapolated values are given for 2000 and 2002. Electron neutrino limits are given for
$\beta$-decay
diamonds) and SN
1987A (green diamonds), for \nmu{} as squares and \ntau as triangles. As can be seen, the
proposed measurement
of $m_{\nmu}$ at
the g-2 \expe{} would result in the largest factor obtained. The mass scale corresponds
to
eV (\nel), keV
(\nmu) and MeV (\ntau) respectively.}
\label{pic:pdg}   
\end{center}
\end{figure}

\section{Summary}
All direct searches for a non-vanishing neutrino mass are currently only resulting
in upper limits. The improvement on the obtained limits of the different \neu{} flavours
is shown in Fig. \ref{pic:pdg}. Nevertheless, three evidences exist which will be
investigated in more
detail within the next years.


\section*{References}
\begin{thebibliography}{99}
\bibitem{zub98} K. Zuber, \Journal{\PRP}{305}{295}{1998}
\bibitem{lob85} V.M. Lobashev et al., \Journal{\NIMA}{240}{305}{1985}
\bibitem{pic92} A. Picard et al., \Journal{\NIMB}{63}{345}{1992}
\bibitem{wei99} C. Weinheimer et al., \Journal{\PLB}{460}{219}{1999}
\bibitem{lob99} V.M. Lobashev et al., \Journal{\PLB}{460}{227}{1999}
\bibitem{gat99} F. Gatti et al., \Journal{Nature}{397}{137}{1999}
\bibitem{ale99} A. Alessandrello et al., \Journal{\PLB}{457}{253}{1999}
\bibitem{dep99} D. Deptuck, private communication
\bibitem{meu98} P. Meunier, \Journal{\NPBP}{66}{207}{1998}
\bibitem{ass96} K. Assamagan et al., \Journal{\PRD}{53}{6065}{1996}
\bibitem{cus99} P. Cushman , K. Jungmann, private communication
\bibitem{raf99} G. Raffelt, Preprint hep-ph/9903472, to appear in ARNPS Vol.49
\bibitem{pas97} L. Passalacqua, \Journal{\NPBP}{55C}{435}{1997}
\bibitem{bar98} R. Barate et al., \Journal{\EPC}{2}{395}{1998}
\bibitem{lee77} B.W. Lee, R.E. Shrock, \Journal{\PRD}{16}{1444}{1977}
\bibitem{mar77} W.J. Marciano, A.I. Sanda, \Journal{\PLB}{67}{303}{1977}
\bibitem{pal92}P.B. Pal, \Journal{\IJMP}{A7}{5387}{1992}
\bibitem{kra90} D. Krakauer et al., \Journal{\PLB}{252}{171}{1990}
\bibitem{abe87} K. Abe et al., \Journal{\PRL}{58}{636}{1987}
\bibitem{kim88} C.S. Kim, W.J. Marciano, \Journal{\PRD}{37}{1368}{1988}
\bibitem{ams97}C. Amsler et al., \Journal{\NIMA}{396}{115}{1997}
\bibitem{bed97} A.G. Beda et al., \Journal{\PAN}{61}{66}{1998}
\bibitem{bar96} I. Barabanov et al., \Journal{\ASP}{5}{159}{1996}
\bibitem{cle98} B.T. Cleveland et al., \Journal{\APJ}{496}{505}{1998}
\bibitem{gal99} W. Hampel et al., \Journal{\PLB}{447}{127}{1999}
\bibitem{abd97} J.N. Abdurashitov et al., Proc. 4th Int. Solar Neutrino Conference,
Heidelberg 1997
\bibitem{bah98} J.N. Bahcall, S. Basu, M.H. Pinnsonneault, \Journal{\PLB}{433}{1}{1998}
\bibitem{bog99}J. Boger et al., Preprint nucl-ex/9910016
\bibitem{borex} Borexino, Proposal LNGS 1991, G. Alimonti et al.,
\Journal{\ASP}{8}{141}{1998}
\bibitem{rag97}R. S. Raghavan, \Journal{\PRL}{78}{3618}{1997}, LENS Collaboration,
Letter of
Intent subm. to INFN/LNGS 1999
\bibitem{tao97} C. Tao, Proc. 4th Int. Solar Neutrino Conference,
Heidelberg 1997
\bibitem{ban95} S.R. Bandler et al., \Journal{\PRL}{74}{3169}{1995}
\bibitem{viu99} J.L. Vuilleumier, talk presented at TAUP'99, Paris
\end{thebibliography}
\end{document}